\documentclass[aps,twocolumn]{revtex4}
\usepackage[amssymb]{SIunits}
\usepackage{SIunits,graphicx,float,amsmath,color}

\begin{document}
\title{Multiple states in highly turbulent Taylor-Couette flow}
\author{Sander G. Huisman${}^*$}
\author{Roeland C. A. van der Veen\footnote{S. Huisman and R. van der Veen contributed equally to this work.}}
\author{Chao Sun}
\email{c.sun@utwente.nl}
\author{Detlef Lohse}
\email{d.lohse@utwente.nl}
\affiliation{Department of Applied Physics and J. M. Burgers Centre for Fluid Dynamics, University of Twente, P.O. Box 217, 7500 AE Enschede, The Netherlands}
\date{\today}

\begin{abstract} 
The ubiquity of turbulent flows in nature and technology makes it of utmost importance to fundamentally understand turbulence. Kolmogorov's 1941 paradigm suggests that for strongly turbulent flows with many degrees of freedom and its large fluctuations, there would only be \emph{one} turbulent state as the large fluctuations would explore the entire higher-dimensional phase space. Here we report the first conclusive evidence of multiple turbulent states for large Reynolds number $\text{Re}=\mathcal{O}(10^6)$ (Taylor number $\text{Ta}=\mathcal{O}(10^{12})$) Taylor-Couette flow in the regime of ultimate turbulence, by probing the phase space spanned by the rotation rates of the inner and outer cylinder. The manifestation of multiple turbulent states is exemplified by providing combined global torque and local velocity measurements. This result verifies the notion that bifurcations can occur in high-dimensional flows (\textit{i.e.} very large $\text{Re}$) and questions Kolmogorov's paradigm. 
\end{abstract}

\maketitle
\section*{Introduction}
For macroscopic flows of water or air the Reynolds number is much larger than unity; the standard type of flow is therefore turbulent. A typical Reynolds number for a person walking is already $\mathcal{O}(10^5)$. For large airplanes the Reynolds number is $\mathcal{O}(10^9)$, for atmospheric currents $\mathcal{O}(10^{10})$, and for ocean currents $\mathcal{O}(10^{11})$. Reynolds numbers are even larger for astrophysical problems~\cite{annurevbalbus}. It is not possible to achieve these large Reynolds numbers in a lab-environment nor is it accessible by direct numerical simulations (DNS). To extrapolate data from $\text{Re}=\mathcal{O}(10^6)$ to the scales of our atmosphere or the ocean we must bridge $4$--$5$ decades in Reynolds number, and even more for astrophysical applications~\cite{ji2013angular}. While scaling laws exist that can predict the rough magnitude of these flows, they are rendered impractical if there is a flow-transition from a turbulent state at lower $\text{Re}$ to another turbulent state at higher $\text{Re}$, or if multiple turbulent states can coexist at the same $\text{Re}$. To extrapolate to large scales we need to know whether there are transitions and whether multiple states can coexist in high Reynolds number flows. As an answer to this question, Kolmogorov's more than 70-year-old paradigm states that for large Reynolds numbers, flows would become `featureless' due to the fact that the highly-dimensional phase space is explored in its entirety due to the large fluctuations of said flows~\cite{kol41,kol41b}. 

For Rayleigh-B\'enard convection at low Rayleigh number (laminar-type boundary layers, $\text{Ra} < 10^{14}$), continuous switching between two different roll states, with different heat transfer properties, was found~\cite{xi2008,poel11,wei13}. In this case, the turbulent fluctuations were large enough to overcome trapping in one turbulent state. In the case of high Rayleigh number (turbulent boundary layers), no multiple states have been observed in a single setup; only when the boundary conditions were changed one could trigger a transition to a different state~\cite{RBchimney,gro11}. For von K\'arm\'an flow multiple turbulent states were found when driving it with impellors with curved blades~\cite{Ravelet-PRL2004,Ravelet-JFM2008,Cortet-PRL2010}. These studies revealed the spontaneous symmetry-breaking and turbulent bifurcations in highly turbulent von K\'arm\'an flow up to $\text{Re}=10^6$. In spherical-Couette flow Zimmerman \textit{et~al.}~\cite{zimmerman2011} observed \textit{spontaneous} switching between two turbulent states at fixed rotation rates. The presence of coherent structures at high Reynolds numbers in closed systems suggests that Kolmogorov's hypothesis~\cite{kol41,kol41b} is incomplete~\cite{Ravelet-PRL2004,Ravelet-JFM2008,Cortet-PRL2010,zimmerman2011} and might need revisiting in order to apply to these flow systems. For Taylor-Couette multiple states are only observed for low $\text{Re}$ (see \textit{e.g.} ref.~\cite{and86})---the so-called classical regime, where the bulk is laminar or turbulent but the boundary layers are still of laminar type. Around $\text{Re}\sim 10^4$~\cite{ost12,ostilla2014a} the system transitions into the ultimate state~\cite{kra62,lathrop1992,gro11}, in which the boundary layers are also turbulent~\cite{hui13}, and where new scaling-laws of the response parameters set in~\cite{rav10,gil11,pao11,gil12,hui12,ost12,brauckmann2013,merbold2013}. Historically, this ultimate regime was defined based on the scaling properties of the flow~\cite{kra62}. Consequently, it does not necessarily exclude the existence of multiple turbulent states. To correctly extrapolate to much higher $\text{Re}$ it is crucial to know the characteristics of the turbulent state and the existence of other such states. In ref.~\cite{Gollub1979} it was shown that for increasing $\text{Re}$ the waves on top of the Taylor-vortices become increasingly complex until only turbulent Taylor-vortices are left. Lewis \textit{et al.}~\cite{lewis1999} came to the same conclusion by plotting the velocity power spectra for increasing Re up to $5\cdot10^5$, and observed that the peaks gradually decrease in amplitude. They noted, though, that turbulent Taylor-vortices remained. On the other hand, the findings of Lathrop \textit{et al.}~\cite{lathrop1992} suggest that the Taylor-vortices are \emph{not} present for Reynolds numbers beyond $1.2\cdot 10^5$.

We will demonstrate that roll structures remain for Taylor-Couette flow even in the ultimate regime up to at least $\text{Re}=\mathcal{O}(10^6)$ and show that multiple states are even possible far beyond the transition into this ultimate regime.

\begin{figure}[h!t]
    \centering
		\includegraphics{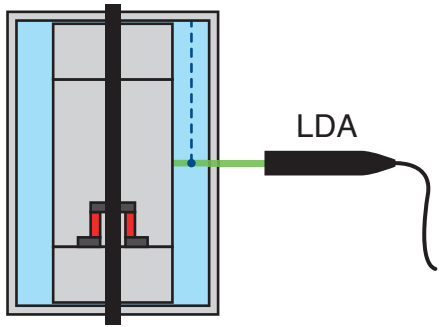}
		\caption{\textbf{Experimental apparatus.} Schematic of the cross section of the T${}^3$C~\cite{gilt3c}: the apparatus has been outfitted with a new coaxial torque transducer (shown in red), see the Methods section. The azimuthal velocity is probed at the middle of the gap and in the top half of the apparatus using laser Doppler anemometry (LDA).}
		\label{fig:setup}
\end{figure}

\begin{figure}[h!t]
    \centering
		\includegraphics{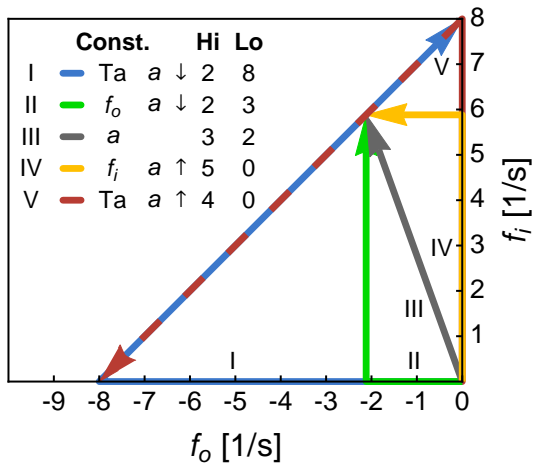}
		\caption{\textbf{Phase space of the trajectories I to V.} The arrows indicate the direction in which the phase space is probed. The legend shows which parameter is kept constant, whether the parameter $a=-f_o/f_i$ is going down ($a\downarrow$) or up ($a\uparrow$), and the number of times it went to a high (Hi) or a low (Lo) state. All trajectories go through (or end at) $f_i = \unit{5.882}{\hertz}$, $f_o = \unit{-2.118}{\hertz}$ ($a=0.36$, and $\text{Ta}  \approx 10^{12}$ or $\text{Re} \approx 10^6$).}
		\label{fig:phasespace}
\end{figure}

\section*{Results}
\textbf{System parameters.} For TC flow~\cite{gil11,pao11,gil12,hui12,ost12,brauckmann2013,merbold2013} and using the analogy of TC flow with Rayleigh-B\'enard (RB) convection~\cite{gro00} it was found that the Taylor number $\text{Ta}=\frac 14 (\frac{1+\eta}{2 \sqrt{\eta}})^4 (r_o - r_i)^2 (r_i + r_o)^2 (\omega_i-\omega_o)^2/\nu^2$ is a very well suited parameter to describe the driving of the system~\cite{eck07b}. Here $\omega_{i,o} = 2\pi f_{i,o}$ are the angular rotation rates and $\nu$ the kinematic viscosity. The response of the system is the torque required to sustain constant angular velocity or a `Nusselt' number~\cite{eck07b} $\text{Nu}_\omega = \tau/\tau_\text{lam}$, which is the angular velocity flux nondimensionalised with the flux of the laminar, non-vortical, flow. This Nusselt number scales approximately as $\text{Nu}_\omega \propto \text{Ta}^{0.38}$~\cite{gil11,pao11} around $\text{Ta}=10^{12}$, which is interpreted as $\text{Nu}_\omega \propto \text{Ta}^{1/2} \cdot \text{Log-corrections}$~\cite{kra62,gro11}. We now find with our new sensor with improved accuracy that the exponent is closer to $0.40$ (results not shown), which is still consistent with the aforementioned interpretation. The Twente Turbulent Taylor-Couette facility ($\text{T}^3\text{C}$)~\cite{gilt3c} was used for the experiments, see fig.~\ref{fig:setup} and the Methods section for more details.

\textbf{Explored phase space.} First we follow trajectory I and V shown in the parameter space of fig.~\ref{fig:phasespace}, which have as a characteristic that $f_i-f_o = \unit{8}{\hertz}$ is kept constant (except for the initial and the final part) and is equivalent to approximately $\text{Ta}=10^{12}$ or $\text{Re}=(\omega_i r_i - \omega_o r_o)(r_o - r_i)/\nu=\mathcal{O}(10^6)$. While traversing the trajectory we continuously measure the torque, scanning over $a$ in one experiment. This is in contrast to experiments that were performed before, where the torque was measured by performing separate ramps of constant $a=-f_o/f_i$~\cite{lathrop1992,gil11,pao11,gil12,merbold2013}. We slowly follow a trajectory in phase space, such as to be in a statistically quasi-steady state the entire time~\cite{gil11}. The temperature variation within the system is \unit{0.04}{\kelvin}, the variation during each measurement is \unit{0.3}{\kelvin}, and the mean temperature for each run is between \unit{19}{\celsius} and \unit{24}{\celsius}. The Taylor number depends on viscosity and thus temperature; we therefore remove the main temperature dependence by compensating the Nusselt number with $\text{Ta}^{0.4}$, because $\text{Nu}_\omega \propto \text{Ta}^{0.4}$ in the present parameter regime. This approach has been followed before, see \textit{e.g.}~\cite{gil11,pao11}.  

\begin{figure}[H]
	{\centering{
		\includegraphics{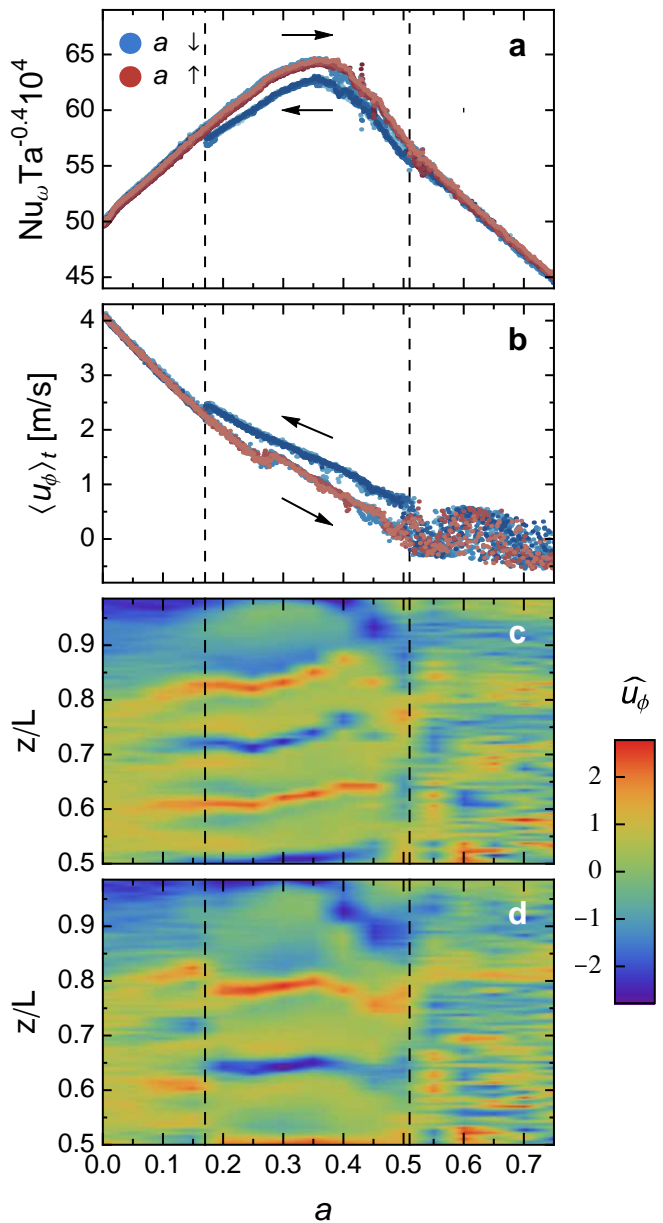}
		\caption{\textbf{Global torque and local velocity at different rotation ratios.} \textbf{(a)} Compensated $\text{Nu}_{\omega}$ as a function of $a$. Experiments following trajectories I and V are colored in blue and red, respectively. Experiments following trajectory I either go into a high or a low state for $0.17<a <0.51$, while experiments of trajectory V are always in the high state. The stability of two different turbulent states is clearly revealed. \textbf{(b)} Azimuthal velocity measured at $r=(r_i+r_o)/2$ and $z/L=0.5$ as a function of $a$ for trajectories I and V. Same colors as in fig.~(a). For the local velocity we also see that the system bifurcates when following trajectory I around $a=0.51$, either choosing the high or low state for $0.17<a <0.51$. \textbf{(c)} and \textbf{(d)} show axial scans of the standardized angular velocity for varying $a$ following trajectories I and V, respectively. We see the presence of 4 rolls in the top half of the system in fig.~(c), while in fig.~(d) only 3 rolls are present for $0.17<a <0.51$. For $a<0.17$ and $a>0.51$ the system is in the same state, regardless of the trajectory followed in phase space. $\widehat{u_\phi} = (u_\phi - \left \langle u_\phi \right \rangle_z)/\sigma_a(u_\phi)$, where $\sigma_a$ is the standard deviation of $u_\phi$ for each $a$. The torque of trajectory V is $2.5\%$ larger than for trajectory I at $a=0.36$.}
		\label{fig:overview}
	}}
\end{figure}

\begin{figure}[h!t]
	\centering
		\includegraphics{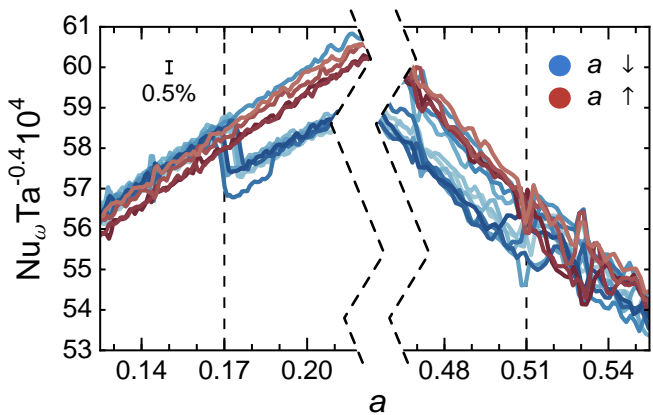}
		\caption{\textbf{Close up view of figure \ref{fig:overview}a.} The trajectories I ($a\downarrow$) that are in the low state sharply transition to the high state around $a=0.17$. Trajectories I start to transition around $a=0.51$, though the process seems more gradual, and sometimes stay in the high state. Trajectories V ($a\uparrow$) never transition and are always in the high state. The error bar, shown in the top-left of the figure, is based on the accuracy of the torque sensor in the system.}
		\label{fig:zoom}
\end{figure}

\textbf{Global torque and local velocity.} As can be seen in figure \ref{fig:overview}, for increasing $a$ (red shades, trajectory V), the torque is continuous and shows a peak around $a=0.36$, as found before~\cite{gil11,pao11,gil12,merbold2013}. For trajectory I the torque is found to be the same as trajectory V for $a<0.17$ and $a>0.51$, however for $0.17<a<0.51$ the torque is found to be different. For decreasing $a$ the system is able to enter another state around $a=0.51$ which is characterised by a lower torque (from here on called `low state'), around $a=0.17$ the system sharply jumps back to a higher torque state (`high state'), see also the close up view in figure \ref{fig:zoom}. We have repeated these experiments in order to see how sharp this transition is, and to see in which state the system is, see fig.~\ref{fig:phasespace}. For trajectory V we observe that it always goes into the high state, while for the reverse trajectory I the system goes to the low state (for $0.17<a<0.51$) with a high probability (8 out of 10).

To verify that the high and low torque states originate from two different physical states, we measure the azimuthal velocity at half-height $z=L/2$ and center of the gap $r=(r_i+r_o)/2$, while the system is let to move along trajectories I and V in phase space, see fig.~\ref{fig:overview}b. As in fig.~\ref{fig:overview}a, it is found that the local velocity inside the system bifurcates and that two states are possible. The presence of multiple states in a local measurement (azimuthal velocity) and at the same time a global measurement (torque) provides convincing evidence that the system can indeed be in different turbulent states, despite the very high Taylor number of $\mathcal{O}(10^{12})$ (ultimate regime). 

\textbf{Flow structure.} In order to further characterize the turbulent state of the system, we perform axial scans of the azimuthal velocity in the top half of our apparatus for several $a$ for both the high and the low state, see figures \ref{fig:overview}c and \ref{fig:overview}d. For each $a$ the azimuthal velocity is standardized (zero mean, unit standard deviation) and color-coded. Fig.~\ref{fig:overview}c shows the local velocity for the high state (trajectory V), and shows the presence of 5 large minima/maxima for $a\leq 0.45$; a clear characteristic of 4 turbulent Taylor vortices. For $a\geq 0.5$ the state of the system is less clear, and the system appears to jump between states (without a well-defined $a$-dependence), as seen in the local velocity in fig.~\ref{fig:overview}b. This behavior looks similar to what was found in RB convection in the classical turbulent regime~\cite{wei13}. However, the mechanism is different due to the higher turnover time scale of the TC system, and the presence of the additional control parameter $a$ in TC, with which we can control the transition. Furthermore, the observed transition is different from Ref.~\cite{zimmerman2011} where \textit{spontaneous} switching back and forth between two states was found, based on global and local measurements. Fig.~\ref{fig:overview}d shows the same switching behavior as the high state of fig.~\ref{fig:overview}c for $a$ outside $[0.17,0.51]$. For $0.17<a<0.51$ it is found that the azimuthal velocity has 4 large minima/maxima, which is the signature of 3 turbulent Taylor vortices (in the top half of the system).

\begin{figure*}[ht!]
	\centering
	\includegraphics{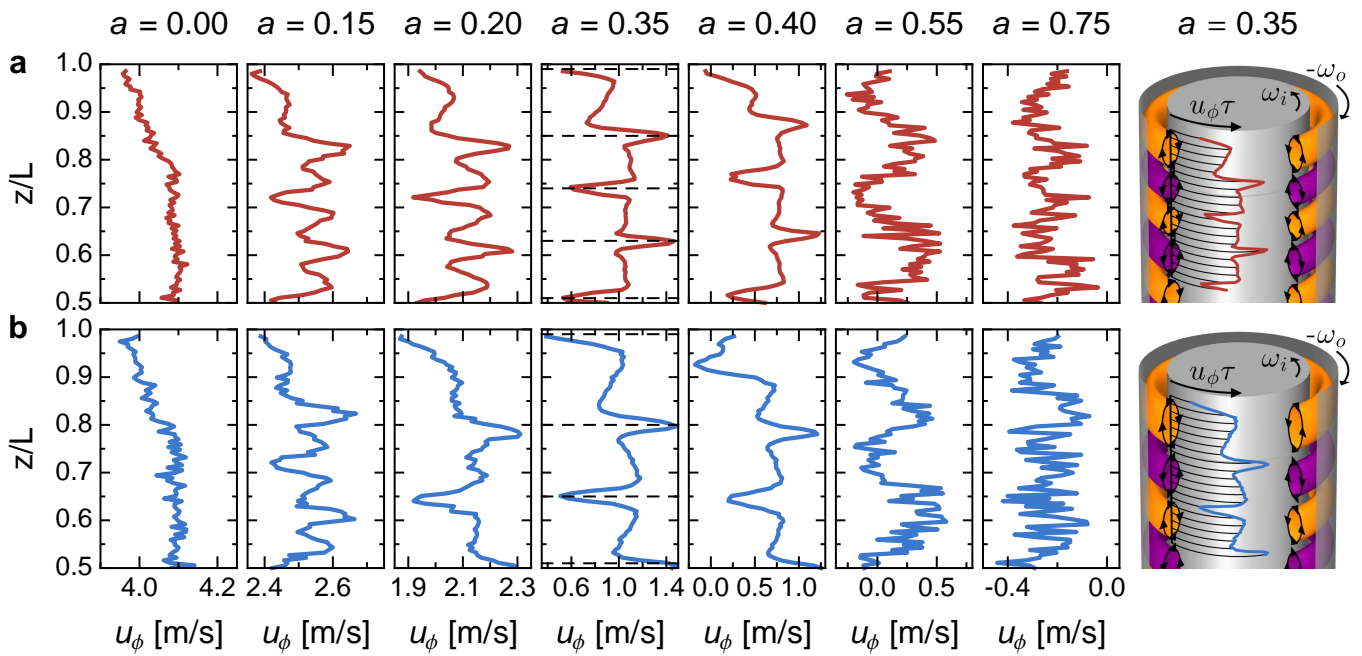}
	\caption{\textbf{Flow structures at different rotation ratios.} Azimuthal velocity as a function of height $z/L$ for various $a$. Subset of the data shown in figures \ref{fig:overview}c and \ref{fig:overview}d. \textbf{(a)} Profiles following trajectory V. \textbf{(b)} Profiles following trajectory I. For $a\in\{0.00,0.15,0.55,0.75\}$ the velocity profiles are the same for both trajectories. For intermediate values ($a\in\{0.20,0.35,0.40\}$) the profiles are different; trajectory V (red) shows the emergence of 4 rolls in the top half of the system, while trajectory I (blue) shows only 3 rolls in the top half of the system. For the case of maximum torque ($a=0.35$), dashed lines are drawn to indicate the boundaries between the rolls. For $a=0.15$ the system shows the high state (with 4 rolls in the top half). This structure seems to fade once the system is pushed towards $a=0$, see also figures \ref{fig:overview}(c) and \ref{fig:overview}(d). On the right we show a schematical overview of the rolls and the angular velocity profile for $a=0.35$ in 3D, $\tau=\unit{0.2}{\second}$.}
	\label{fig:profiles}
\end{figure*}

In addition, we provide, for selected $a$, the angular velocity profiles as lines in fig.~\ref{fig:profiles}. The angular profiles for trajectories I and V are identical (within experimental and statistical error) for $a$ outside $[0.17, 0.51]$. For $a\in[0.17, 0.51]$ the profiles are different and show distinguishing features of 3 or 4 rolls (6 and 8 rolls in the entire setup), see also the schematics on the right of fig.~\ref{fig:profiles}. At the boundary of the last (top) roll and the penultimate roll high velocity fluid from the inner cylinder is advected towards the middle, increasing the velocity at the center of the gap. Similarly, at the boundary of the penultimate and the antepenultimate roll the low velocity fluid from the outer cylinder is advected inwards, decreasing the velocity at the center of the gap. The corresponding average aspect ratio of the vortices is $1.95$ (3 rolls) and $1.46$ (4 rolls), which is consistent with previous studies (see \textit{e.g.} fig.~2.5 in ref.~\cite{chouippe-phd}). Close inspection of fig.~\ref{fig:overview}c and fig.~\ref{fig:profiles}a shows that the roll around $z/L=0.5$ slightly drifts upwards for $0.3<a<0.45$ (compared to $a<0.3$), which could be a signature of symmetry-breaking; the top half of the system behaves slightly differently from the bottom half. This is an explanation of the anomaly in the local angular velocity for trajectory V around $a\approx0.28$. Such symmetry-breaking behavior has also been found in von K\'arm\'an flow~\cite{burguette2013}.

From our findings of trajectory I and V (constant $\text{Ta}$) we find that the maximum torque is at $a=0.36$ ($f_i=\unit{5.882}{\hertz}$, $f_o=\unit{-2.118}{\hertz}$). We now look at other trajectories reaching this maximum but keeping either $f_o$, $a$, or $f_i$ constant, see trajectories II--IV in fig.~\ref{fig:phasespace}. For constant $f_o$ and constant $a$ (trajectories II and III) the system was found to have the ability to go in either the high or the low state as well, see the legend of fig.~\ref{fig:phasespace}. Like trajectory V, trajectory IV (constant $f_i$) was found to be characterised by a high state of the system. It seems that if $a=0.36$ is approached from below (trajectories IV and V) the flow does not bifurcate. However, the system does bifurcate when $a=0.36$ is approached from the top, or if $a$ is kept constant but the driving strength is increased.

\section*{Discussion}
We have shown that Taylor-Couette flow displays flow structure even for a very high Taylor number of roughly $10^{12}$ ($\text{Re}=\mathcal{O}(10^6)$), which is beyond Reynolds numbers for which large-scale structures were believed to vanish in Kolmogorov's picture. In addition, we found that the system is hysteretic and can be in multiple stable turbulent states for the same driving parameters. The multiple states are simultaneously measured globally and locally by performing torque and LDA measurements. It was found that multiple states can occur for rotation ratios $0.17<a<0.51$. For $0\leq a<0.17$ there is only a single stable state, though we cannot exclude that other trajectories in phase space might trigger multiple states in this region. For $a>0.51$ the system does not posses a state with a clear roll structure. Presently, a theoretical understanding of the values $a=0.17$, $a=0.51$ and the sharp and smooth behaviour of the jumps around those $a$ is lacking. Finally we note that the presented experiments, performed in the T${}^3$C facility, will be challenging to simulate in DNS in the foreseen future: simulations of such high $\text{Ta}$ are difficult, especially for $\Gamma=\mathcal{O}(10)$. The present work highlights the importance of the coherent structures and their selectability in highly turbulent flows, which demand continued effort and investigation. The question of whether or not these structures survive for even larger Reynolds numbers, remains open but is important for understanding the myriad of large-scale flows in nature.

\section*{Methods}
The $\text{T}^3\text{C}$ \cite{gilt3c} has an inner cylinder with an outer radius of $r_i = \unit{200}{\milli \meter}$, a transparent outer cylinder with inner radius $r_o = \unit{279}{\milli \meter}$, and a height of $L = \unit{927}{\milli \meter}$, giving a radius ratio of $\eta =r_i/r_o = 0.716$ and an aspect ratio of $\Gamma = L/(r_o - r_i) = 11.7$. The top and bottom caps rotate along with the outer cylinder. The apparatus was filled with water and actively cooled to keep the temperature constant. 

The torque is measured on the middle section of the inner cylinder using a new co-axial torque transducer (Honeywell \texttt{2404-2K}, maximum capacity of \unit{225}{\newton \meter}), with improved accuracy compared to our former load cell \cite{gil11}. 

The middle section of the inner cylinder with height $z/L=0.578$ does not cover all the rolls, therefore the exact mean torque value over the entire inner cylinder could be different. It is, however, unlikely that this difference takes away the `jumping' behavior mentioned before.

The azimuthal velocity is obtained by laser Doppler anemometry (LDA), see fig.~\ref{fig:setup}. The laser beams go through the outer cylinder and are focused in the middle of the gap, see fig.~\ref{fig:setup} The water is seeded with \unit{5}{\micro \meter} diameter polyamide tracer particles (Dantec) with a maximum Stokes number of $\text{St}=\tau_p/\eta_K=0.004 \ll1$. Curvature effects of the outer cylinder to our LDA system are accounted for by numerically ray-tracing the LDA-beams~\cite{hui12lda}.


\section*{Acknowledgments}
We would like to thank M. Bos, G.W. Bruggert, and G. Mentink for their technical support and G. Ahlers and S. Grossmann for discussions. This work was financially supported by an European Research Council (ERC) Advanced Grant, the Simon Stevin Prize of the Technology Foundation STW of The Netherlands, and the European High-Performance Infrastructures in Turbulence (EuHIT).

\section*{Author contributions}
S.G.H. and R.C.A.v.d.V. designed and performed the experiments and analyzed the data. C.S. and D.L. supervised the project. All authors discussed the physics and contributed to the writing of the manuscript. 

\end{document}